 \documentstyle[twoside,fleqn,espcrc2,epsfig]{article}

\newcommand{\AmS}{{\protect\the\textfont2
  A\kern-.1667em\lower.5ex\hbox{M}\kern-.125emS}}

\title{Lepton asymmetries from neutrino oscillations}

\author{R. R. Volkas\address{School of Physics, 
        Research Centre for High Energy Physics, \\ 
        The University of Melbourne, Victoria 3010, Australia}%
        \thanks{Invited talk at Neutrino 2000. 
                Work supported by the Australian Research Council.}}

\begin{document}

\begin{abstract}
Reasonably large relic neutrino asymmetries can be generated by active-sterile
neutrino oscillations. After briefly discussing possible applications, I describe
the Quantum Kinetic Equation formalism used to compute the asymmetry growth curves.
I then show how the basic features of these curves can be understood on the
basis of the adiabatic limit approximation in the collision dominated epoch, and
the pure MSW effect at lower temperatures.
\vspace{1pc}
\end{abstract}

\maketitle

\section{BRIEF OVERVIEW}

\begin{figure*}[htb]
\hspace{1cm}\epsfig{file=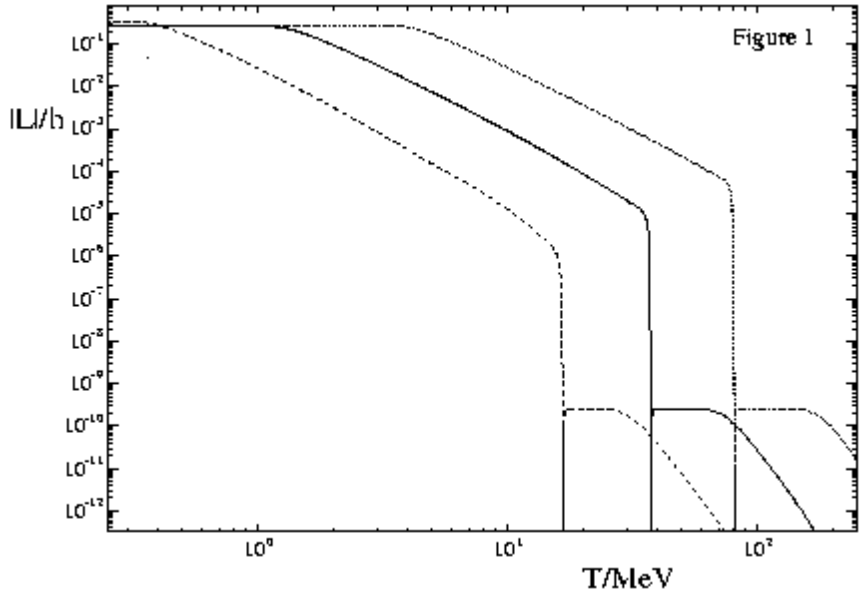, width=12cm}
\vspace{-1cm}
\caption{Examples of lepton asymmetry growth curves driven by $\nu_{\tau}
\leftrightarrow \nu_s$ and the corresponding antineutrino oscillations. The
mixing angle is selected to be $\sin^2 2\theta = 10^{-8}$. The three curves
correspond to $\Delta m^2 = -0.5,\ -50,\ -5000$ eV$^2$, reading left to
right. 
This figure is taken from Ref.\protect\cite{footastropart}.}
\label{fig1}
\end{figure*}

The $\alpha$ flavour relic neutrino asymmetry $L_{\nu_{\alpha}}$ is defined by
\begin{equation}
L_{\nu_{\alpha}} = \frac{ n_{\nu_{\alpha}} - n_{\overline{\nu}_{\alpha}} }{ n_{\gamma} }
\end{equation}
where $n_{\psi}$ is the number density of species $\psi$, and $\alpha = e,\mu,\tau$.
Reasonably large values for $L_{\nu_{\alpha}}$ will be generated by $\nu_{\alpha}
\leftrightarrow \nu_s$ and $\overline{\nu}_{\alpha} \leftrightarrow \overline{\nu}_s$
oscillations (where $s$ denotes a sterile species), in the temperature range $1$ MeV to 
$10$'s
of MeV, provided that $\Delta m^2 < 0$ and the vacuum mixing angle $\theta_0$ is small
\cite{ftv,fv1,fv2,footastropart}.
(Note that $|\Delta m^2|$ must be sufficiently large to induce significant
asymmetry growth prior to neutrino decoupling at $1$ MeV. We focus on this case.)
By convention, $\Delta m^2 < 0$ means that the predominantly sterile mass eigenstate
is lighter than the predominantly active eigenstate. The asymmetry growth is seeded by
the $CP$ asymmetric background plasma (baryon/electron asymmetry), and it is 
driven by runaway positive
feedback conditions that prevail for a short time below a critical temperature $T_c$.
Figure \ref{fig1} shows $|L_{\nu_{\alpha}}|$ evolution 
for three different vacuum oscillation parameter
choices \cite{footastropart}.

It is noteworthy that this mechanism requires light sterile neutrinos. These
hypothetical degrees of freedom are interesting for other reasons also: as a
simple explanation for two-fold maximal mixing (via the mirror matter \cite{mirror}
or pseudo-Dirac schemes \cite{pd}), and in order to reconcile the solar \cite{suzuki}, 
atmospheric \cite{sobel}
and LSND anomalies \cite{lsnd}.

Large neutrino asymmetries have found three potential applications. The first
is the suppression of sterile neutrino production \cite{ftv,fv1,footastropart,dibari1}. 
Recall that the matter-affected
mixing angle for neutrinos is given by
\begin{equation}
\sin^2 2\theta_m = \frac{ s^2 }{ s^2 + (b - a - c)^2 }
\end{equation}
where $s \equiv \sin 2\theta_0$ and $c \equiv \cos 2\theta_0$. The functions
$a$ and $b$ contribute to the effective potential \cite{msw}
\begin{equation}
V_{\alpha} = \frac{ \Delta m^2 }{ 2p } (-a + b)
\end{equation}
where $p$ is the neutrino momentum or energy and
\begin{eqnarray}
a & = & - \frac{2 p \sqrt{2}G_F n_{\gamma}L}{\Delta m^2},
\nonumber\\
b & = & - \frac{2 \sqrt{2} p^2\,G_{F}\,n_{\gamma}}{\Delta m^2}
\frac{A_{\alpha}T}{m^{2}_{W}}.
\end{eqnarray}
The quantity $G_F$ is the Fermi constant, $m_W$ is the $W$ boson mass, $A_e \simeq 17$,
$A_{\mu,\tau} \simeq 4.9$ and the {\it effective total lepton number} $L$ is
\begin{equation}
L \equiv L_{\nu_\alpha} + L_{\nu_e} + L_{\nu_\mu}
+ L_{\nu_\tau} + \eta, 
\end{equation}
where $|\eta| \sim 5 \times 10^{-10}$ is a small term due to the asymmetry of the 
electrons and nucleons.
The $a$ term is the Wolfenstein effective potential, while $b$ is a finite temperature
correction that is very important in the early universe context. When $V_{\alpha}$ is large,
the matter-affected mixing angle is small and sterile neutrino production is
suppressed. At high $T$, the $b$ term alone is sufficient for this purpose, but at
low $T$ a large enough $L$ is required.

The second application requires an electron-like asymmetry $L_{\nu_e}$ of the
appropriate magnitude. Primordial light element abundances depend on
the expansion rate of the universe during Big Bang Nucleosynthesis (BBN) as well as the
reaction rates for $\overline{\nu}_e p \leftrightarrow e^{+} n$ and $\nu_e n
\leftrightarrow e^{-} p$. The latter are affected by $L_{\nu_e}$, with a positive asymmetry
lowering $n/p$ and thus also the Helium abundance. Now, the Helium yield increases with
the baryon density $\eta_B$. 
A positive $L_{\nu_e}$ thus allows one to obtain an acceptable $^4$He
mass fraction with a higher baryon density than would otherwise be the case \cite{fv2,bfv}.
Interestingly, the recent Boomerang/Maxima cosmic microwave background anisotropy
measurements prefer a relatively high $\eta_B$ \cite{cmbr}. 
This is a hint that $L_{\nu_e} > 0$
should be taken seriously for phenomenological reasons.

The final application is related to the second. Recall that the effective potential
depends on the linear combination $L$ which contains the baryon asymmetry $\eta$.
Because of this, neutrino asymmetry generation depends on the properties of
the baryonic component of the plasma. If the latter were to be spatially inhomogeneous,
then it could seed inhomogeneous asymmetry creation \cite{dibari2}. 
There could be a domain structure,
whereby the sign of $L_{\nu_e}$ flips between spatial cells. This will
become especially interesting if future observations 
uncover firm evidence for inhomogeneous light element abundances.

Another interesting issue is the possibility of rapid oscillations in the asymmetry
immediately after $T_c$ \cite{ftv,shi,ropa}. 
Recent work has shown that there are definitely
no $L_{\nu_{\alpha}}$
oscillations for a large region of oscillation parameter space \cite{ropa}. 
The remaining
region requires significant computational effort, and has not been properly 
explored. There are indications for rapid oscillations, but numerical error cannot be
definitively ruled out as their cause. On the theoretical front, it is strongly 
suspected that non-adiabatic effects should cause oscillations \cite{fv1,new1}, 
so it would not
surprise if numerical work eventually confirms same (but only for a certain
region of parameter space).

\section{QUANTUM KINETIC EQUATIONS}

The dynamical variable of interest is the 1-body reduced density matrix $\rho$
for the $\nu_{\alpha}/\nu_s$ subsystem. There is a corresponding density matrix
for the antineutrinos. These functions depend on time $t$, or equivalently
temperature $T$, and momentum $y \equiv p/T$.
We will employ the decomposition
\begin{equation}
\rho \equiv \frac{1}{2} (P_0 + \vec{P}\cdot\vec{\sigma}).
\end{equation}
The diagonal entries are conveniently normalised distribution functions
$f_\alpha$ and $f_s$ for $\nu_{\alpha}$ and $\nu_s$, respectively:
\begin{equation}
f_{\alpha}  =  \frac{1}{2}(P_0 + P_z) f_{eq}^0, \quad
f_s  =  \frac{1}{2}(P_0 - P_z) f_{eq}^0.
\end{equation}
The off-diagonal entries $P_{x,y}$ are the coherences.

The Quantum Kinetic Equations (QKEs) are \cite{stod}
\begin{eqnarray}\label{eq:b1}
& {\partial \vec{P} \over \partial t}  = 
\vec{V} \times \vec{P} - D(P_x  {\bf \hat x} 
+ P_y {\bf \hat y})+ {\partial P_0 \over \partial t}\, {\bf \hat z},& \\
& {\partial P_0 \over \partial t}  \equiv  R \simeq 
\Gamma(y)\left\{{f_{eq} \over f^0_{eq}} - {1\over 2}(P_0 + P_z)
\right\}.&
\label{yd}
\end{eqnarray}
The damping or decoherence function $D$ is given by
$D = \frac{\Gamma}{2}$, 
with $\Gamma$ being 
the total collision rate of the weak 
eigenstate neutrino of momentum $y$ with the background 
plasma,
$\Gamma (y) \simeq \kappa_{\alpha}\,G_F^2T^5\,y$,
where $\kappa_{e} \simeq 1.27$ and 
$\kappa_{\mu}=\kappa_{\tau} \simeq 0.92$ 
(for $m_e \stackrel{<}{\sim} T \stackrel{<}{\sim} m_\mu$). 

The function $f_{eq}$ is the Fermi-Dirac distribution,  
\begin{equation}
f_{eq}(y) \equiv {1 \over 1 + e^{y- \tilde{\mu}_{\alpha}}}.
\end{equation}
where $\tilde{\mu}_{\alpha} \equiv \mu_{\nu_\alpha}/T$ is a
dimensionless chemical potential, and $f^0_{eq}$ is
the $\mu_{\nu_{\alpha}}=0$ case.

The quantity $\vec{V}$ is given by
\begin{equation} 
{\bf V} = \beta {\bf \hat x} + \lambda {\bf \hat z},
\end{equation}
where
\begin{equation}
\beta = \frac{\sin 2\theta_0}{\ell_0},\quad
\lambda = -\frac{\cos2\theta_0}{\ell_0} + V_{\alpha},
\label{sf}
\end{equation}
and $\ell_0 \equiv 2yT/\Delta m^2$ is the vacuum
oscillation length.

The $\vec{V} \times \vec{P}$ term describes coherent nonlinear MSW
evolution, the $D$ entry represents collisional decoherence, while
$R$ quantifies repopulation of the $\nu_{\alpha}$ distribution from the background
plasma.

An overbar will designate corresponding antineutrino functions. The
kinetic equations are identical except for the substitution $L \to - L$.

The QKEs together with $\alpha + s$ lepton number conservation supply
\begin{equation}
\frac{ d L_{\nu_{\alpha}} }{ dt } = \frac{T^3}{2n_{\gamma}} \int_0^{\infty}
\beta (P_y - \overline{P}_y) \frac{y^2f_{eq}^0}{2\pi^2} dy
\label{Lev}
\end{equation}
as the asymmetry evolution equation. This is a redundant equation, but numerically
very useful in removing the problem of taking the difference
of two large numbers to obtain the asymmetry. It is nonlinear, because $\lambda$
and $\overline{\lambda}$ depend on $L_{\nu_{\alpha}}$.

\section{ANALYTICAL AND PHYSICAL UNDERSTANDING OF ASYMMETRY GROWTH}

The growth curves in Fig.\ref{fig1} were obtained by numerically solving the QKEs
in conjunction with Eq.(\ref{Lev}). In this section I review how the
QKE ``black box'' may be opened and the
basic features of these graphs analytically and physically understood.
This will provide a partial reply to the claims of Ref.\cite{dhps}. A more
complete riposte can be found in Ref.\cite{comment}.

\subsection{Collision dominated epoch}

The collision rate $\Gamma$ is proportional to $T^5$, so at high enough
temperatures collisions dominate the evolution.
In this epoch it is legitimate to take $R \simeq 0$, because $f_{\alpha}$
does not depart from equilibrium form very strongly. See Ref.\cite{new2} for a
discussion. The QKEs then reduce to
\begin{equation}
{\partial {\bf P}\over \partial t} \simeq {\cal K}{\bf P}
=\left( 
\begin{array}{ccc}
-D   & -\lambda & 0    \\ 
 \lambda & -D   & -\beta \\ 
 0   & \beta  & 0 
\end{array} \right) \left( \begin{array}{c}
P_x \\ P_y \\ P_z \end{array} \right).
\end{equation}
In the instantaneous diagonal basis ${\bf Q} = {\cal U}{\bf P}$
this equation becomes
\begin{equation}
\frac{\partial {\bf Q}}{\partial t} = {\cal K}_{d} {\bf Q} - {\cal U}
\frac{\partial {\cal U}^{-1}}{\partial t} {\bf Q},
\label{Q}
\end{equation}
where ${\cal K}_d$ is the diagonal matrix of eigenvalues $k_{1,2,3}$ of ${\cal K}$.
In the adiabatic limit we set ${\partial {\cal U}^{-1}}/{\partial t} = 0$ and
Eq.(\ref{Q}) may be formally solved to yield \cite{new1,bvw}
\begin{equation}
{\bf P}(t) \simeq {\cal U}^{-1}(t)
e^{\int_0^t {\cal K}_d dt'}
{\cal U}(0){\cal P}(0).
\label{formal}
\end{equation}

Under most circumstances, the eigenvalues take the form \cite{new1}
\begin{equation}
k_1 = k_2^* = - d + i \omega,\quad k_3 = - \frac{\beta^2 D}{d^2 + \omega^2},
\end{equation}
with
\begin{equation}
d = D + \frac{k_3}{2},\quad \omega^2 = \lambda^2 + \beta^2 + k_3 D + \frac{3}{4}k_3^2.
\end{equation}
In the high $T$ collision dominated regime
\begin{equation}
d \simeq D,\quad \omega \simeq \sqrt{\beta^2 + \lambda^2}
\end{equation}
and
\begin{equation}
k_3 \simeq - \frac{\beta^2 D}{D^2 + \lambda^2 + \beta^2} \ll D,
\end{equation}
except at the very centre of the resonance. For full technical details
concerning the resonance regime see Ref.\cite{new1}. When collisions dominate,
$\exp(\int k_{1,2} dt') \to 0$, and Eq.(\ref{formal}) yields
\begin{equation}
P_y(t) \simeq \frac{k_3}{\beta} P_z(t),
\end{equation}
so that Eq.(\ref{Lev}) becomes
\begin{equation}
\frac{ d L_{\nu_{\alpha}} }{ dt } = \frac{T^3}{2n_{\gamma}} \int_0^{\infty}
(k_3 P_z - \overline{k}_3 \overline{P}_z) \frac{y^2f_{eq}^0}{2\pi^2} dy.
\label{Lad1}
\end{equation}
One also obtains a sterile neutrino production equation which I will not display.
These equations involve the distribution functions only -- the coherences
have been eliminated -- so we have a {\it Boltzmann limit}. The root of this
lies with the collisional damping by $-d$ of the oscillatory 
behaviour driven by $\omega$.

Substituting for the various functions on the righthand side of Eq.(\ref{Lad1})
one finally obtains, to leading order \cite{ftv,fv1,bvw},
\begin{displaymath}
\frac{ d L_{\nu_{\alpha}} }{ dt } \simeq \frac{s^2 T^3}{4\pi^2n_{\gamma}} \times \nonumber
\end{displaymath}
\begin{equation}
\int_0^{\infty} dy
\frac{ \Gamma a (c - b) (f_{\alpha} + f_{\overline{\alpha}} - f_s - f_{\overline{s}}) y^2}
{[x + (c - b + a)^2][x + (c - b - a)^2]} ,
\label{Lad2}
\end{equation}
where $x \equiv s^2 + (D\ell_0)^2$. This has been called the ``static approximation''.
There are correction terms also which I will
not discuss. If $\Delta m^2 > 0$, then it is easy to see 
$d L_{\nu_{\alpha}}/dt \sim - L$ and so $L_{\nu_{\alpha}}$ evolves so as to
drive the effective asymmetry $L$ to zero. The $\Delta m^2 < 0$ case is, however,
quite different.
The function $b$ is then always positive, decreasing as $T^5$, and the sign of
the asymmetry derivative changes from negative to
positive at a critical temperature $T_c$ due to the 
evolving net effect of the $(c-b)$ factor in the integrand. Immediately
after $T_c$, a brief spurt of exponential growth occurs because now
$d L_{\nu_{\alpha}}/dt \sim + L$.

\begin{figure*}[htb]
\hspace{2cm}\psfig{file=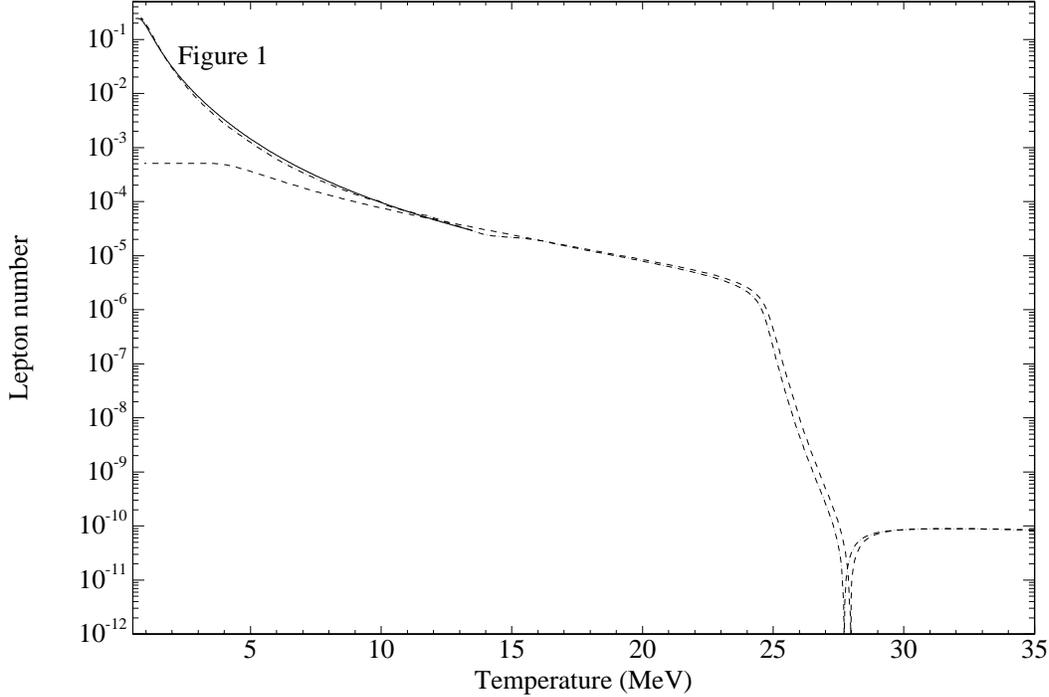,width=12cm,angle=-90}
\vspace{-4cm}
\caption{Asymmetry growth curves from the QKEs (dash-dotted line), 
the static approxmation (dashed line)
and undamped adiabatic MSW transitions (solid line) as presented
in Ref.\protect\cite{fv2} for $\Delta m^2 = -10$ eV$^2$ and
$\sin^2 2\theta_0 = 10^{-9}$.}
\label{fig2}
\end{figure*}

The dashed line in
Fig.\ref{fig2}, taken from Ref.\cite{fv2}, shows the asymmetry growth curve obtain from
Eq.(\ref{Lad2}). The dash-dotted curve displays asymmetry evolution as obtained
from the full QKEs for the same oscillation parameter choice. The explosive
growth at $T_c$ is well approximated by Eq.(\ref{Lad2}), demonstrating that
collision dominated adiabatic evolution is the dominant physical effect
during this epoch. At lower $T$, the magnitude of the asymmetry is seriously
underestimated. New physics has taken over \cite{fv2}.

\subsection{MSW dominated epoch}

We begin by observing that Eq.(\ref{Lad1}) reduces to $dL_{\nu_{\alpha}}/dt = 0$
when the collision rate $\Gamma$ vanishes. This helps to explain why the
approximate evolution equation reviewed above underestimates asymmetry growth
at lower temperatures. When collisions are unimportant, the QKEs describe
pure (non-linear) matter-affected evolution, so we now turn to a study
of the MSW effect.

Putting $D = \Gamma/2 = 0$ in the adiabatic evolution matrix ${\cal K}$ 
produces the eigenvalues
\begin{equation}
k_1 = k_2^* = i \sqrt{\beta^2 + \lambda^2},\quad k_3 = 0.
\end{equation}
There is no damping so
the evolution is oscillatory, with $|k_{1,2}|$ being the matter-affected
oscillation frequency for a neutrino of momentum $y$. The conditions for
the validity of Eqs.(\ref{Lad1}) and (\ref{Lad2}) no longer hold \cite{new1,bvw}.

\begin{figure*}[htb]
\hspace{2cm}\psfig{file=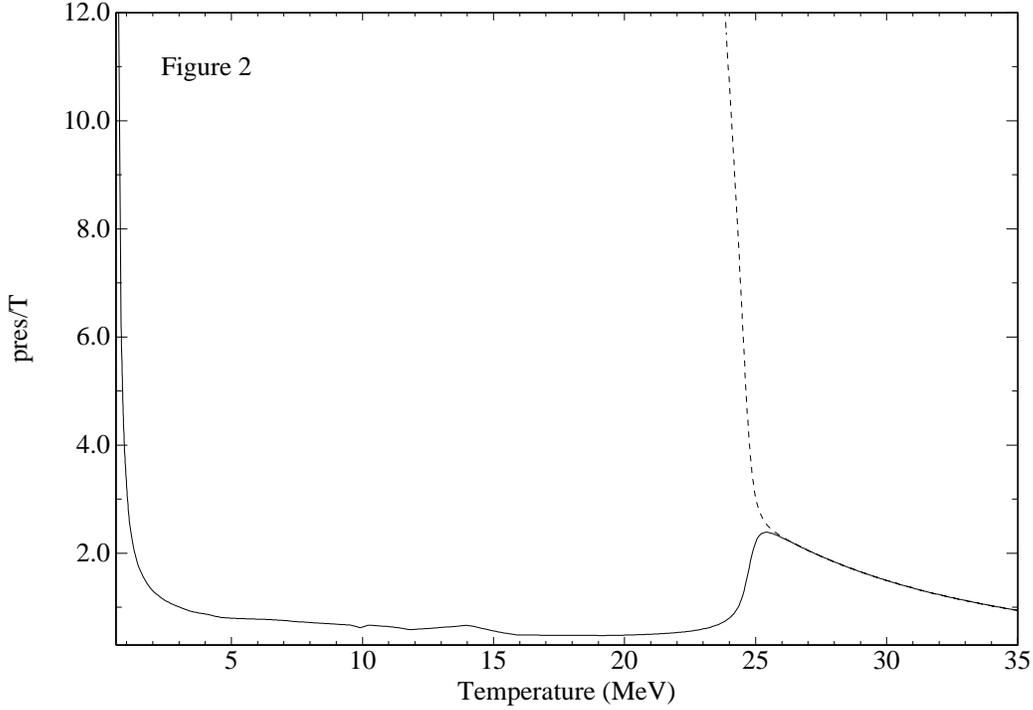,width=12cm,angle=-90}
\vspace{-4cm}
\caption{The separation of the neutrino and antineutrino resonance momenta
as the asymmetry grows. The oscillation parameters are as for Fig.\ref{fig2}.
This figure first appeared in Ref.\protect\cite{fv2}.}
\label{fig3}
\end{figure*}

Furthermore, the condition of the plasma is now different because of
the significant neutrino asymmetry, leading to a separation of the
neutrino and antineutrino resonance momenta $y^r$ \cite{fv2}. 
Writing $a \equiv \xi_a y$
and $b \equiv \xi_b y^2$, the MSW resonance conditions $c - b \mp a = 0$
imply that
\begin{equation}
y^{r}_{\nu} = \frac{ \xi_a + \sqrt{\xi_a^2 + 4\xi_b}}{2\xi_b}
\end{equation}
and
\begin{equation}
y^{r}_{\overline{\nu}} = \frac{ - \xi_a + \sqrt{\xi_a^2 + 4\xi_b}}{2\xi_b}.
\end{equation}
If $L > 0$, then $\xi_a$ is positive and $y^{r}_{\nu}$ quickly moves
to the high-momentum tail of the distribution. The partial cancellation
between the terms in the numerator of $y^{r}_{\overline{\nu}}$ keeps
the antineutrino resonance within the body of the distribution. Doing
the algebra with $\xi_b \ll \xi_a$ one obtains
\begin{equation}
y^{r}_{\overline{\nu}} \simeq \frac{\pi^2 |\Delta m^2|}{8\sqrt{2} \zeta(3) G_F}
\frac{1}{L_{\nu_{\alpha}} T^4},
\label{LT4}
\end{equation}
where we have taken $L \simeq 2 L_{\nu_{\alpha}}$. The resonance momentum
evolution \cite{fv2} is shown in Fig.\ref{fig3}.

The
separation of the resonances means that the MSW effect has very
asymmetric (pardon the pun) consequences for neutrinos and antineutrinos:
neutrino oscillation are strongly suppressed, whereas $\overline{\nu}_{\alpha}$'s
get MSW converted into sterile states. This keeps the positive $L$ growing \cite{fv2}.
(The final sign of $L$, but not its magnitude, 
depends on the high $T$ initial neutrino chemical potentials and so cannot
be predicted.)
The MSW effect is of course automatically taken care of by the QKEs. But in
order to isolate its influence, a more physical approach is useful. We will
stay with the $L > 0$ case. For small resonance widths and $100\%$ (adiabatic)
conversion, the rate of change of the asymmetry is proportional to the
speed with which the antineutrino resonance moves through the distribution \cite{fv2}:
\begin{displaymath}
\frac{ d L_{\nu_{\alpha}} }{ dT } = - \frac{T^3}{2\pi^2 n_{\gamma}}\,
[f_{\overline{\nu}_{\alpha}}(y^{r}_{\overline{\nu}}) - f_{\overline{\nu}_{s}}
(y^{r}_{\overline{\nu}})]\,
(y^{r}_{\overline{\nu}})^2\, 
\frac{ d y^{r}_{\overline{\nu}} }{ dT } 
\end{displaymath}
\begin{equation}
\equiv - X  
\frac{ d y^{r}_{\overline{\nu}} }{ dT }
\label{preseq}
\end{equation}
Using Eq.\ (\ref{LT4}) one then obtains
\begin{equation}
\frac{ d L_{\nu_{\alpha}} }{ dT } = - \frac{4}{T}
\frac{X y^{r}_{\overline{\nu}}}
{ 1 + \frac{X y^{r}_{\overline{\nu}}}{L_{\nu_{\alpha}}} }
\label{LMSW}
\end{equation}
as a non-linear evolution equation for the asymmetry, for the case
where $\frac{d y^{r}_{\overline{\nu}}}{d T} < 0$. We employ
instantaneous repopulation before neutrino decoupling, 
so $f_{\overline{\alpha}}$ is
understood to be of Fermi-Dirac form with the appropriate
(evolving) chemical potential. After chemical and kinetic decoupling
one has to handle repopulation in a more complicated way: see Refs.\cite{fv2,bfv}
for further explanations.

The solid line in Fig.\ref{fig2} shows the result of numerically integrating
Eq.(\ref{LMSW}), beginning at $T \simeq T_c/2$. The agreement with
the QKE result is excellent, which shows that the MSW effect has taken
over by $T \simeq T_c/2$ \cite{fv2}. Furthermore, when $L_{\nu_{\alpha}} \ll 1$,
Eq.(\ref{LMSW}) reduces to
\begin{equation}
\frac{ d L_{\nu_{\alpha}} }{ dT } \simeq - \frac{4 L_{\nu_{\alpha}}}{T},
\end{equation}
which immediately implies that $L_{\nu_{\alpha}} \sim T^{-4}$. The power
law behaviour displayed by the QKE solution is thus easily understood
as a consequence of undamped adiabatic MSW transitions \cite{fv2}.

As the asymmetry continues  to grow, the antineutrino resonance
eventually
moves out of the body into the high momentum end of the
distribution. The asymmetry becomes frozen at some final steady
state value, whose approximate magnitude can be easily understood
by integrating the Fermi--Dirac distribution 
from $y \sim 0$ to $y = \infty$ \cite{fv2}:
\begin{displaymath}
L^{\rm final}_{\nu_{\alpha}} \sim
\frac{1}{4\zeta(3)}
\left(\frac{T_{\nu_{\alpha}}}{T_{\gamma}}\right)^3 \int_0^{\infty}
\frac{ y^2 dy}{1 + e^y}
\end{displaymath}
\begin{equation}
\label{freeze}
= \frac{3}{8}\left(\frac{T_{\nu_{\alpha}}}{T_{\gamma}}\right)^3.
\end{equation}
The temperature ratio takes care of reheating due to $e^{+}e^{-}$
annihilations at $T \simeq m_e \simeq 0.5$ MeV. Equation
(\ref{freeze}) gives the approximate magnitude only, because the
distribution changes with time as the asymmetry is created.
Numerically, the final values found by incorporating proper
thermalisation effects are quite close to this estimate.

\section{CONCLUSIONS}

Active-sterile neutrino oscillations will produce a final steady state
asymmetry of order $3/8$ provided the oscillation parameters are in the
appropriate range. Applications include the suppression of sterile neutrino 
production prior to BBN and the alteration of the primoridal Helium abundance due
to a $\nu_e$ asymmetry. Inhomogeneous baryogenesis can seed lepton domain
formation leading to inhomogeneous BBN. The asymmetry growth curves
obtained from brute force numerical solution of the QKEs have been
understood analytically and physically.

\end{document}